\documentclass{IEEEtran}
\usepackage{cite}
\usepackage{color, soul}
\usepackage{amsmath,amssymb,amsfonts}
\usepackage{algorithmic}
\usepackage{graphicx}
\usepackage{textcomp}
\usepackage{hyperref}
\def\BibTeX{{\rm B\kern-.05em{\sc i\kern-.025em b}\kern-.08em
    T\kern-.1667em\lower.7ex\hbox{E}\kern-.125emX}}
    
\begin{document}
\title{Enabling and Enforcing Social Distancing Measures using Smart City and ITS Infrastructures: \\ A COVID-19 Use Case}
\author{\centering
    \IEEEauthorblockN{\centering
         Maanak Gupta\IEEEauthorrefmark{1}, 
        Mahmoud Abdelsalam\IEEEauthorrefmark{2}, 
        Sudip Mittal\IEEEauthorrefmark{3} 
        \\
    }

    \IEEEauthorblockA{
    \IEEEauthorrefmark{1}
        \textit{Tennessee Technological University}, Cookeville, TN, USA \\
        mgupta@tntech.edu \\
    }
     \IEEEauthorblockA{
    \IEEEauthorrefmark{2}
        \textit{Manhattan College}, Riverdale, NY, USA \\
        mabdelsalam01@manhattan.edu \\
    }
    \IEEEauthorblockA{
    \IEEEauthorrefmark{3}
        \textit{University of North Carolina Wilmington}, Wilmington, NC, USA  \\
        mittals@uncw.edu \\
    }
}

\maketitle

\begin{abstract}
Internet of Things is a revolutionary domain that has the caliber to impact our lives and bring significant changes to the world. Several IoT applications have been envisioned to facilitate data driven and smart application for the user. Smart City and Intelligent Transportation System (ITS) offer a futuristic vision of smart, secure and safe experience to the end user, and at the same time efficiently manage the sparse resources and optimize the efficiency of city operations. However, outbreaks and pandemics like COVID-19 have revealed limitations of the existing deployments, therefore, architecture, applications and technology systems need to be developed for swift and timely enforcement of guidelines, rules and government orders to contain such future outbreaks. This work outlines novel architecture, potential use-cases and some future directions in developing such applications using Smart City and ITS. 
\end{abstract}


\section{Introduction}
\label{sec:introduction}
The COVID-19 pandemic has engulfed the world with a large portion of the population infected. These staggering numbers highlight various opportunities in existing technologies and infrastructure to contain the spread of this highly contagious virus. The impact of Coronavirus is both global and unpredictable. Also, the supply chain shock it is causing will most definitely and substantially cut into the worldwide manufacturing revenue of US \$15 trillion currently forecasts for 2020 by global tech market advisory firms~\cite{ABI_research}.

The lack of an approved medicine to cure this disease presses the need for prevention and mitigation mechanisms to minimize the spread. \textit{Social distancing} measures, including country wide lock-downs, travel prohibition, quarantining hot-spots, and limiting customers at essential businesses, are slowly restricting the spread of the virus. It is paramount, in such critical situations, to implement swift mechanisms, and various agencies work in coordination to limit the spread of the disease. It is expected that broad range of situational intelligence and automated targeted response is needed throughout the community to ensure the safety of people and assets, with an aim to lower the fatalities and minimize impact on the economy. However, even in the cities where strict social distancing guidelines were issued to limited movement, it has been noticed the technology can play a role in enforcing such restrictions. In the United States~\cite{smartcitiesdive}, New York, San Francisco, New Orleans, and Philadelphia ranked as the most vulnerable specifically in terms of health risks. These large cities are vulnerable due to high population density and higher use of public transportation, according to the report. Also, about 61\% percent of New York City residents commute using public transportation compared to less than 0.1\% of residents from Amarillo, TX, which ranked as one of the least vulnerable cities in the United States. 

It is clearly evident, with the surge in the number of cases for COVID-19, that our cities are not prepared for fending off such pandemics and outbreaks, assisting in tracking the suspects, limiting movement of people to vulnerable neighborhoods, directing and monitoring traffic, or even pushing alerts to commuters. There are some initiatives worldwide which were deployed on ad-hoc basis to fight this pandemic. Countries and cities deployed drones with loudspeakers to communicate rules and government mandated precautions with residents. Wearable devices \cite{sphcc} have been mandated for the population to enforce quarantine measures, whereas telecommunication data has been used to monitor the large gathering of people. Corporations are  developing applications for contact tracing \cite{contact-tracing,singapore-contact} which can effectively be enabled to trigger an alert if someone gets in contact with newly diagnosed COVID-19 patients. 

Currently many countries around the world are approaching the idea of deploying ``smart'' cities, including USA, Netherlands, China, and Japan, to say the least. For instance, Amsterdam's latest model of \textit{SmartCity 3.0}~\cite{ams-smartcity} motivates active participation of citizens and private organizations along with the government in creating and expanding smart city solutions. This initiative includes research, projects and data sharing in areas such as infrastructure and technology, smart energy and water, and Intelligent Transportation System (ITS). Another prime example is Smart America~\cite{smartamerica}, a White House Presidential Innovation Fellow project with the goal to research smart city related topics such as ITS, Smart Manufacturing, Healthcare, Smart Energy, and Disaster Response. Although many cities are taking the initiative of turning into smart cities, there are many challenges that need to be addressed.

To be effectively prepared for future outbreaks and enforce protocols, it is expected that smart cities and ITS technology will host a range of data driven services together with deployed sensors to not only help in enforcing community wide social distancing measure but also assist in early detection of such outbreaks. This article focuses on proposing a novel architecture along with several use-cases which can be developed to create a smart city and ITS inspired data driven system which can help to enable and enforce community distancing measures during pandemics and high impact low frequency (HILF) events. We also highlight challenges and future opportunities to empower future smart cities and ITS to cater such circumstances.
\section{Architecture}
Figure 1 shows a holistic overview of the smart city and ITS architecture. The proposed conceptual model specifically caters to the requirements and support applications which are envisioned to enforce social distancing and community measures in pandemics and high impact low frequency (HILF) events like COVID-19. In simple terms, this architecture provides the common basis for planners and engineers to conceive, design and implement system together with concerns relevant to large number of stakeholders to offer services, alerts and data driven applications. This service oriented architecture is scalable to offer applications in wide geographic location offering quality of service (QoS). 

Overall, the architecture consists of physical devices, including road-side sensors, smart traffic lights and connected cars, having the ability to record real time data and exchange messages with nearby entities and upload relevant information to central cloud facilities for processing. At the same time, such smart physical devices can also replicate edge computes to offer the capability for real time low latency communication supported by ITS such as vehicle-to-vehicle (V2V), vehicle-to-infrastructure (V2I) or vehicle-to-everything (V2X). The flexibility of the architecture to adapt to various use-case requirements can be achieved by having a hybrid edge cloud supported model which enable dynamic real time needs, and at the same time offer infinite computation on data captured using central cloud infrastructures. Basic Safety Messages (BSMs) can also be used to enable V2X communication with information pertaining to an event, location or even severity. It is also viable to create \textit{digital twins} for various smart devices to offer a virtual counterpart for each physical object. These twins, based on the data collected in the real environment, can provide insights to improve the operations, increase efficiency and discover issues.

The participating entities including vehicles, law enforcement, drones, parking sensors or other roadside units (RSUs) must enroll with a central authority to receive certificates and ensure trustworthiness of messages exchanged among entities. The communication technologies which can enable the messages exchange can include cellular LTE, WiFi, 5G or Dedicated Short Range Communication (DSRC). In addition, Message Queuing Telemetry Transport (MQTT) messaging protocol has been widely used to support physical entity to cloud service providers like Amazon Web Services (AWS) or Microsoft Azure. This architecture will need multiple technologies and communication protocols to cater different use cases. As shown in the figure, the communication (with dashed lines) can happen among different physical entities, using cloud supported services, edge assisted messaging or even peer to peer model. Such infrastructures and supported technologies will need long term investment and partnership among public and private entities to enable hundreds of smart devices interaction with each other.  

The architecture further focuses on offering data sharing platform across interested parties which can help speed up reaction times and also enforce community measures in critical situations. By real time data sharing with law enforcement, city authorities or other stakeholders, it is expected that more efficient and automated response mechanisms can be implemented. For example, in order to ensure social distancing and limiting community gatherings, drones can be used to collect information about the number of individuals and messages can be sent to the near by traffic signals to divert traffic to other location. At the same time, automatic disinfectant and fogging can be deployed in certain locations based number of existing cases, vulnerability or large gatherings. Parking availability can be shared in real time with the cloud services and registered individuals can be allotted a time to ensure minimal human interface. These applications (as more discussed in the next section) allow cities to gain insights and identity problems with low response time, and at the same time enable efficient resource planning in times with high resource demand in most affected areas. With continuous monitoring, data capturing and real time communication, an efficient and more productive safer environment can be created.

\begin{figure}[t]
\centering
\includegraphics[scale=0.45]{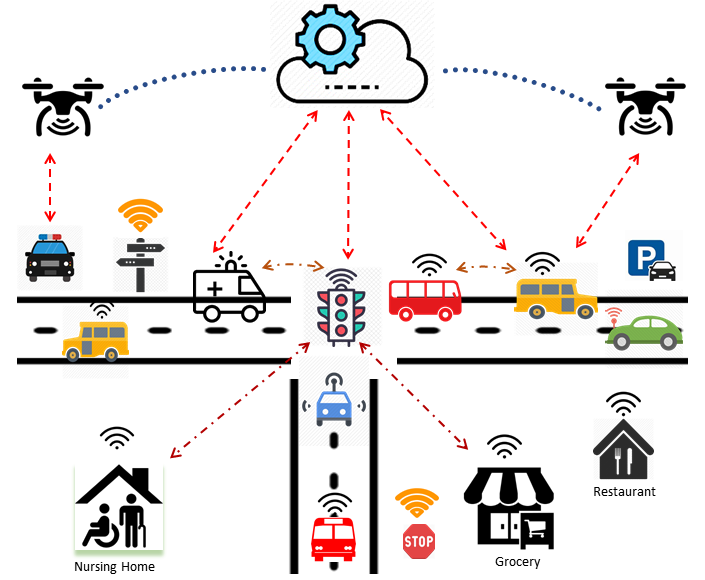}
\caption{Smart City and ITS Architecture} 
\label{fig-arch}
\end{figure}

\section{Usecases}

In this section, we will discuss some important usecases that describe how smart city infrastructure and various Intelligent Transportation systems technologies can be used to enforce and enable social distancing measures in a swift and timely manner, together with data driven AI applications. 

\subsection{Monitoring Large Gatherings}
Monitoring large gatherings during pandemics is essential to regular social distancing. ITS and Smart cities infrastructures can play an important role in such a scenario. During the COVID-19 outbreak, multiple countries have started utilizing drones and other AI assisted technologies to enforce social distancing rules (including lock down rules). For example, state police in Western Australia, Derbyshire police in England, the local police in Italy, and local police in New Jersey~\cite{us-drones} plan to utilize drones to enforce social distancing. The drones can be used to monitor areas like, recreational parks, beaches, and public transits, ensuring people comply with the most recent forged gathering rules. Based on the severity of infection in a particular place, more or less stringent rules can be used. This can be done by using body heat temperature sensors to detect the number of people present at one particular place. Upon detecting a higher than the allowed number, drones can instruct the people to evacuate the place due to a higher chance of getting infected. It can also be used to detect proximity of people to enforce distancing rules. If a particular area is known to be a hot infectious spot, a complete lock down can be enforced. 

Drones can also help in informing people about the newest enforced rules. By taking advantage of ITS and smart cities infrastructures, different rules can be enforced in different locations. However, ongoing change of social distancing rules can be hard to keep up with, especially during an outbreak in which people are in a state of stress. Using drones with attached loud speakers to inform the people at a particular location can be very useful. This will help ensure that people are aware of the recent enforcement rules to prevent unintended break of rules. These drones can also communicate with automated disinfectant spray vehicles to remotely monitor there effectiveness and direct to needed communities.
%
As such having a systematic way of incorporating drones as part of ITS infrastructure, in particular, and smart cities infrastructure, in general, can be very useful.

\subsection{Smart Parking}

Smart parking~\cite{khanna2016iot} can play an important role in enforcing social distancing. In smart parking, sensors and other AI based techniques are used to determine whether a particular spot on the street or in a parking garage is occupied or vacant. Using such information, different restrictions can be added on the number of vehicles parked at a particular location. For example, a parking space at a grocery store might allow more cars to be parked at one particular time than a senior home which is more critical due to the occupancy of elderly people who are more prone to COVID-19. Changing the parking rules at different locations, each with its own parking rules to alleviate gathering, can be challenging. Specially, since such rules can be updated everyday and informing law enforcement agents of these updates can be a daunting task. Smart parking can automatically enforce these rules while informing the vehicles through relayed messages.



\subsection{Re-Routing Traffic to Reduce Footprint}



ITS and smart city infrastructures can be useful in creating on the fly designated, high speed routes to critical and vulnerable locations like hospitals to supply resources like ventilators, Personal Protective Equipment (PPEs), etc. With the shortage of such supplies due to high volume of patients infected, many states are pooling all the surgical resources available in state hospitals and assigning them on per need basis. In such a scenario, real time count of the supplies can be maintained in hospitals, that can automatically initiate a message to the central pool location. Once initiated, designated enforcement vehicles can be loaded with emergency supplies and fastest routes can be created on the fly with smart traffic lights, and roadside units to provide free passage. At the same time V2V and V2I messaging can be used among vehicles and infrastructures to ensure fastest delivery. 

It is also possible to create AI assisted systems that direct vehicles to drive through COVID-19 testing locations based on an area and waiting time at different locations. This can be enforced with automated waiting count at various testing areas and sending such data to a central cloud, which will then forward messages to vehicles directing them to different available locations. Such enforcement can also limit movement of visitors in old-age homes or localities with greater density of senior citizen population. More stringent policies can be implemented in states and counties with high rates of health conditions like, obesity, diabetes, hypertension, heart, and kidney disease. Such architecture can also enable the number of delivery drivers from restaurants and grocery stores to allow a certain number of people at a time in any location. 

\textit{E-Passes} can be generated using Radio-frequency identification (RFID) tags that can be placed on a vehicle, to keep track on who is on the road. This enforcement can also help in limiting the number of vehicles on the road at one time. For example, it should be enforced that no one from one household is allowed to visit grocery store more than once in 10 days. If the vehicle registered for a household is detected on road, it can be confiscated, and an automatic ticket/fine can be issued. It is also possible to assign a designated time and date for a particular vehicle belonging to a household. If another vehicle from the same household is detected by roadside sensors without an approved RFID, law enforcement can take policy based actions. Such RFID can be issued at a nearby Department of Motor Vehicles (DMVs) which can also allow only one vehicle from each household to be used for driving. In case of front-line workers like healthcare providers, first responders, or delivery personnel, exemptions can be made on a case by case basis for allowed vehicles. 

As a result of various social distancing policies limiting store/building occupancy, people are waiting outside various businesses like grocery stores and pharmacies. Queues have been formed outside such essential businesses to comply with various maximum headcount policy. In such a scenario, people are still exposed to risk of aerosol infection because of the waiting time outside the stores even though social distancing is followed. Using ITS and smart infrastructures, vehicles can input their destination in the car Global Positioning System (GPS), which can interact with a central command authority to decide if it is feasible and allowed to go to the designated location. Various AI assisted systems can be built to add this functionality for a smart city. A vehicle can be directed to other similar stores with a lower foot count and less wait time. It is also possible to send a message to the car from the store about various designated parking spots available. Law enforcement can automatically be notified in case of violations and automatically issue tickets/fines.

\subsection{ITS, Big Data \& AI}

An ITS collects a lot of data about vehicles and traffic patterns. Specifically, Dedicated Short Range Communications (DSRC) messages have been used in ITS to codify communication. DSRC are one-way or two-way short-range to medium-range wireless communication channels specifically designed for automotive use. These include fields like, latitude, longitude, time, heading angle, speed, acceleration, brake status, steering angle, headlight status, wiper status, external temperature, turn signal status, vehicle length, vehicle width, vehicle mass. Augmented with other data sources which include, ambient temperature, ambient air pressure, traction control status, road conditions, etc.

This data source can serve as a medium to create multiple big data and AI systems that can provide value by enabling and enforcing social distancing measures. One such big data system can be leveraged to find traffic hotspots in a smart city. These traffic hot spots can inform the local government about areas which are frequented by the population and must be monitored regularly. Finding these traffic hotspots is also necessary during the disinfection phase. Multiple cities all over the world have deployed heavy machinery to spray antiseptic solutions along high traffic areas such as, markets, streets, shops, religious and public buildings. Leveraging smart city big data and AI applications will help empirically prioritize specific locations which are proven traffic hotspots. In a futuristic smart city environment, a system can be built to deploy drones to clean, disinfect and monitor these traffic hotspots.

An ITS system can also be used to track vehicles currently on the road. This specific information can be used by the local law enforcement to create policies that enable enforcement of social distancing measures. An example situation for such a case is tracking vehicles which have come from areas that have a large number of COVID-19 cases. Here the DSRC messages or AI based registration plate readers, can be used to read state registration number plates that can inform the local authorities about the origins of a vehicle.


\begin{figure}[t]
\centering
\includegraphics[width=.5\textwidth]{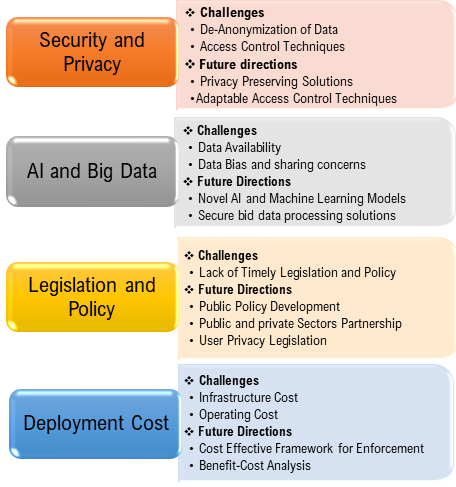}
\caption{Research Challenges and Future Directions.} 
\label{fig-arch}
\end{figure}
\section{Challenges and Future Directions}

Building such a social distancing enforcement system at a smart city level is challenging. Researchers and developers need to keep these in mind as in when they are developing such systems. Some of them have been discussed below.

\textbf{Security and Privacy -} One concern relates to the security and privacy ~\cite{van2016privacy, fries2012meeting} of general citizens who live in smart cities and drive on roads equipped with intelligent transportation systems. While building these AI and big data systems, researchers need to make sure that they de-anonymize the data required. We should also use various access control techniques to ensure the data maintains its confidentiality, availability, and integrity. In the future, these smart cities will generate a large amount of personal data as in when citizens will interact with the smart city. Security and privacy preserving techniques need to be built in parallel as we develop these smart cities. 

\textbf{AI \& Big Data -} Another challenge in creating various big data and AI systems \cite{chauhan2016addressing} is availability and the amount of data needed. Most of the big data and AI applications described above require a large quantity of labeled data. Multiple partnerships must be developed between the private and the public sector to create datasets that can then be leveraged to build these smart city models.

When we develop various AI assisted usecases, special attention needs to be given to keep these models fair and without bias. Techniques that ensure bias free decision computations are being created and need to introduced in various smart city systems. 

\textbf{Legislation and Policy -}
As stated earlier, privacy is one technical concern for social distancing using smart cities and ITS; however, technological means must complement legislation~\cite{ncsl}. Smart city and ITS technologies are rapidly on the rise but legislation and policies are lacking and may fall short for a long period of time. Formal policies for social distancing privacy are necessary and  must address several issues including secure data collection, video surveillance, users/residents consent, and trusted third-parties involved, to say the least. Such legislation must be abided by when deploying smart cities and ITS technologies. As such, research on legislation and policy, including engagement on public policy development debates, is necessary to successfully integrate technical privacy solutions in the smart city and ITS ecosystem.

Not only privacy related legislation is needed, but other policies that encourage innovation, competition, and private investment in social distancing enforcement using smart cities and ITS technologies are essential. Such policies should include incentives and support for partnership between public and private sectors and removal of government barriers. They should also highlight societal benefits of advancements in large-scale social distancing enforcement such as economical growth and saving lives.

\textbf{Deployment Cost -} Deployment cost of smart cities and ITS is massive ranging from infrastructure costs to operating costs. For instance, attaching devices to electric public lighting not only needs municipality collaboration but also isolation from other critical systems which poses a burden on time and effort and, in turn, increases the costs. Another example, battery powered devices (e.g., drones) are limited in the amount of time they operate and the information they can transmit by the battery lifetime. Not to mention, regular maintenance of such devices can be very costly, given they are usually deployed in large numbers. Costs also include operating costs. For instance, NYC operated an Automated Speed Enforcement program in 140 school zones with 180 speed cameras for over two years at a cost of \$69,460,446~\cite{itscosts}. It is worth mentioning that enforcing social distancing may add on to those aforementioned costs due to the need of particular devices and sensors (e.g., drones equipped with body heat sensors for crowd detection).

Even though the deployment costs of smart cities and ITS are massive, they can help reduce costs on other occasions~\cite{lam2017study}. For example, drones monitoring crowds for social distancing can save the expenses of law enforcement officers physically monitoring the streets and allow them to attend to more crucial matters. Additionally, enforcing social distancing in and by itself reduces costs. During the COVID-19 outbreak cutting down the number of patients will substantially reduce the healthcare sector costs. As such, developing a framework for social distance enforcement in an effective, timely and cost-effective manner is an important future research direction.

Another important future research direction includes benefit-cost analysis studies for smart cities and ITS. As an example, ITS traffic re-routing and predictive analysis can help reduce traffic congestion. Not only will it serve in enforcing social distancing rules but also save costs of unnecessary trips to crowded destinations (e.g., grocery stores). These, including aforementioned scenarios, are examples where enforcing social distancing helps in saving money. Analysis on the degree to which enforcing social distancing rules potentially reduces costs is crucial as it incentivizes the government and reduces the economic burdens of applying a large-scale social distancing enforcement.

\section{Conclusion}
COVID-19 outbreak is unprecedented and has disrupted lives of millions of people across the globe. This pandemic has opened several research challenges and opportunities that our community must address to equip itself for the future. The proposed architecture and AI assisted applications discussed in the article can be used to effectively and timely enforce social distancing community measures, and optimize the use of resources in critical situations. This article offers a conceptual overview and serves as a steppingstone to extensive research and deployment of automated data driven technologies in smart city and intelligent transportation systems. For future, we envision to develop these AI driven applications for wider adoption in the community.

\bibliographystyle{unsrt}
\bibliography{bibothe}

\begin{thebibliography}{10}

\bibitem{ABI_research}
{ABI research}.
\newblock URL:
  \url{https://iotbusinessnews.com/2020/03/17/07945-coronavirus-outbreak-reveals-the-weakest-links-in-the-supply-/chain-the-suppliers-supplier/}
  (Accessed on: 04/11/2020).

\bibitem{smartcitiesdive}
{Smart Cities Dive}.
\newblock URL:
  \url{https://www.smartcitiesdive.com/news/clever-real-estate-most-vulnerable-cities-covid-19/575010/}
  (Accessed on: 04/11/2020).

\bibitem{sphcc}
{SPHCC employs IoT tech and wearable sensors to monitor COVID-19 patients}.
\newblock URL:
  \url{https://www.mobihealthnews.com/news/asia-pacific/sphcc-employs-iot-tech-and-wearable-sensors-monitor-covid-19-patients}
  (Accessed on: 04/11/2020).

\bibitem{contact-tracing}
{Apple, Google Bring Covid-19 Contact-Tracing to 3 Billion People}.
\newblock URL:
  \url{https://www.bloomberg.com/news/articles/2020-04-10/apple-google-bring-covid-19-contact-tracing-to-3-billion-people}
  (Accessed on: 04/11/2020).

\bibitem{singapore-contact}
{Coronavirus: Singapore develops smartphone app for efficient contact tracing}.
\newblock URL:
  \url{https://www.straitstimes.com/singapore/coronavirus-singapore-develops-smartphone-app-for-efficient-contact-tracing}
  (Accessed on: 04/11/2020).

\bibitem{ams-smartcity}
{Amsterdam Smart City 3.0}.
\newblock URL:
  \url{https://smartcityhub.com/governance-economy/amsterdam-better-than-smart/}
  (Accessed on: 04/12/2020).

\bibitem{smartamerica}
{Smart America}.
\newblock URL: \url{https://smartamerica.org/} (Accessed on: 04/11/2020).

\bibitem{us-drones}
{Drones in Enforcing Social Distancing}.
\newblock URL:
  \url{https://nymag.com/intelligencer/2020/04/social-distancing-enforcement-drones-arrive-in-the-u-s.html}
  (Accessed on: 04/12/2020).

\bibitem{khanna2016iot}
Abhirup Khanna and Rishi Anand.
\newblock Iot based smart parking system.
\newblock In {\em 2016 International Conference on Internet of Things and
  Applications (IOTA)}, pages 266--270. IEEE, 2016.

\bibitem{van2016privacy}
Liesbet Van~Zoonen.
\newblock Privacy concerns in smart cities.
\newblock {\em Government Information Quarterly}, 33(3):472--480, 2016.

\bibitem{fries2012meeting}
Ryan~N Fries, Mostafa~Reisi Gahrooei, Mashrur Chowdhury, and Alison~J Conway.
\newblock Meeting privacy challenges while advancing intelligent transportation
  systems.
\newblock {\em Transportation Research Part C: Emerging Technologies},
  25:34--45, 2012.

\bibitem{chauhan2016addressing}
Sumedha Chauhan, Neetima Agarwal, and Arpan~Kumar Kar.
\newblock Addressing big data challenges in smart cities: a systematic
  literature review.
\newblock {\em info}, 2016.

\bibitem{ncsl}
{National Conference of State Legislatures}.
\newblock URL:
  \url{https://www.ncsl.org/research/telecommunications-and-information-technology/consumer-data-privacy.aspx}
  (Accessed on: 04/11/2020).

\bibitem{itscosts}
{U.S. Department of Transportation, Office of the Assistant Secretary for
  Research and Technology}.
\newblock URL: \url{https://www.itscosts.its.dot.gov/its/benecost.nsf/CostHome}
  (Accessed on: 04/11/2020).

\bibitem{lam2017study}
Patrick~TI Lam and Wenjing Yang.
\newblock A study of the costs and benefits of smart city projects including
  the scenario of public-private partnerships.
\newblock {\em International Journal of Urban and Civil Engineering},
  11(5):600--605, 2017.

\end{thebibliography}

\end{document}